\documentclass[10pt]{article}

\usepackage[a4paper,margin=2.2cm]{geometry}
\usepackage[T1]{fontenc}
\usepackage{lmodern}
\usepackage{microtype}
\usepackage{amsmath,amssymb,amsthm,mathtools}
\usepackage{enumitem}
\usepackage{xcolor}
\usepackage[hidelinks]{hyperref}
\usepackage{booktabs}

\setlist{nosep}
\newcommand{\Tr}{\operatorname{Tr}}
\newcommand{\id}{\operatorname{id}}
\newcommand{\Down}{\Downarrow}
\newcommand{\Reach}{\operatorname{Reach}}
\newcommand{\kernel}[1]{#1^{\downarrow}}
\newcommand{\bisim}{\sim}
\newcommand{\treq}{\equiv_{\mathrm{tr}}}
\newcommand{\Cexec}{C_{\mathrm{exec}}}
\newcommand{\Crec}{C_{\mathrm{recognize}}}
\newcommand{\nnz}{\operatorname{nnz}}
\newcommand{\width}{\operatorname{width}}
\newcommand{\adm}{\Sigma_{\mathrm{adm}}}
\newcommand{\T}{\mathcal{T}}

\theoremstyle{plain}
\newtheorem{proposition}{Proposition}
\newtheorem{theorem}[proposition]{Theorem}
\newtheorem{lemma}[proposition]{Lemma}

\theoremstyle{definition}
\newtheorem{definition}[proposition]{Definition}

\theoremstyle{remark}

\title{\textbf{Endogenous Interpretation}\\
\large Semantic Deobfuscation as Architecture-Constrained Recovery\\
of Factored Transition Models}
\author{Antonio Nappa\\
\small ACM \\ \texttt{anappa@acm.org}}
\date{September 2026 --- working paper, v2}

\begin{document}
\maketitle

\begin{abstract}
We propose \emph{endogenous interpretation} as a viewpoint on executable
programs: relative to a host substrate, program, interpreter, and emulated
machine are parameters of one transition relation rather than disjoint
semantic objects. Each executable representation carries an \emph{implicit
constraint bias}, the structural restrictions imposed by its instruction
basis, state encoding, control transfers, and finite substrate. Under this
viewpoint a transformation that preserves a chosen observable behaviour does
not remove the computation that produces it; it redistributes that
computation over another state space and encoding, which we call
\emph{semantic diffusion}. \emph{Semantic deobfuscation}---a sub-problem of
reverse engineering, distinct from the recovery of provenance such as types,
names, and intent---is then the recovery of a low-complexity representative
of the behavioural equivalence class under a chosen observation model.

We give two formal instances and keep them separate. For an
\emph{explicitly given} finite-state realization, contraction under branching
bisimilarity is canonical and computable by partition refinement, and this
extends to trace equivalence when both the realization and the contraction
are required to be deterministic; if the analyst admits nondeterministic
contractions, minimal contraction under trace equivalence is PSPACE-hard
again. The tractability boundary is thus set by the observation model and
the hypothesis class, not by the transformation. For \emph{inference from
executions}, where the analyst holds concrete traces rather than the
transition system, we pose deobfuscation as minimum-description-length
recovery of a factored transition model---a dependency hypergraph with local
transition functions---over an architecture-constrained family of
factorizations. The number of nonzeros of the transition tensor is invariant
under all such reshapings; what obfuscation inflates is the factored
description length and the size of the version space of consistent
factorizations. We catalogue the sources of that ambiguity (gauge, coverage,
dependency, role, granularity, level, and observation-model ambiguity), relate
each standard protection to the ambiguity it induces, and state four
independent falsifiable experiments. Factored transition models and MDL
structure recovery are established; the contribution claimed is their
connection to binary semantics, virtualization, and code reuse through an
architecture-aware hypothesis class in which program and interpreter roles
are recovered as part of the structure.
\end{abstract}

\section{Introduction}

\subsection{Endogenous interpretation and implicit constraint bias}

An executable substrate can induce further executable languages from its own
available state transformers. An instruction set, an interpreter, a virtual
machine, or a code-reuse gadget set may each provide a basis from which
programs are formed. Relative to a fixed host, the distinction among
``program,'' ``interpreter,'' and ``emulated machine'' depends on which
parameters of the host's transition relation are held fixed; we call this
\emph{endogenous interpretation}.

Every such representation is constrained. Its instruction vocabulary, state
layout, addressing rules, control transfers, calling conventions, decoder,
and finite implementation restrict the space of realizable descriptions. We
call the resulting structural preference an \emph{implicit constraint bias}.
It is not a heuristic imposed by the analyst; it is carried by the executable
representation, because a transformed program must respect the constraints of
some substrate in order to run. The working conjecture of this note is that
semantic deobfuscation succeeds in practice partly because these constraints
restrict the otherwise enormous space of factorizations the analyst must
consider.

\subsection{Vocabulary: $O$-semantics, provenance, and which gap}

Throughout, ``semantics'' means \emph{observable behaviour under a chosen
observation model $O$}, written $O$-semantics. A transformation is
$O$-preserving if it preserves that behaviour. This is the sense in which
compiler-correctness results are stated \cite{leroy2009}: preservation of a
chosen set of observable traces, not of every internal operational fact.
Ordinary compilers and obfuscators preserve I/O behaviour and do not preserve
intermediate states, evaluation order, or memory layout; we never claim they
preserve ``operational semantics in full.''

Two different ``semantic gaps'' appear in the literature. In virtual-machine
introspection the phrase names the distance between raw guest memory and
operating-system abstractions \cite{garfinkel2003,jain2014}. In decompilation
it names the distance between machine code and source-level meaning. We are
concerned with the latter, which we call the \emph{decompilation gap}.

Within the decompilation gap, two things are lost at once. Compilation and
obfuscation genuinely destroy \emph{provenance}: source identifiers, types,
data-structure boundaries, module structure, invariants, intent. Recovering
provenance is a large part of reverse engineering and is \emph{not} the
subject of this note. What a transformation cannot destroy, if the result is
to execute, is the structure needed to reproduce the $O$-observable
behaviour. We distinguish
\[
  \boxed{\text{provenance loss} \neq \text{$O$-semantic loss},}
\]
and confine the theory to the second.

\subsection{Scope: semantic deobfuscation, not reverse engineering}

Reverse engineering recovers types, protocols, algorithms, vulnerabilities,
design decisions, and intent, most of which are provenance. The problem this
note models is narrower:
\[
  \boxed{
  \text{semantic deobfuscation}=
  \text{recovery of a low-complexity representative of }[P]_O
  \text{ under a chosen }O.
  }
\]
Semantic deobfuscation is a prerequisite for much provenance recovery (one
cannot type what one cannot read), but it is not the whole task, and no
claim here should be read as one about reverse engineering in general.

\subsection{Contributions}

\begin{enumerate}
  \item A vocabulary (endogenous interpretation, implicit constraint bias,
        semantic diffusion, semantic contraction) under which compilation,
        interpreter towers, virtualization, and code-reuse computation appear
        as refactorizations of one host transition relation
        (\S\ref{sec:endogenous}--\ref{sec:diffusion}).
  \item For explicitly given finite-state realizations: canonical contraction
        under branching bisimilarity; its extension to trace equivalence under
        a deterministic hypothesis class; the return of PSPACE-hardness when
        nondeterministic contractions are admitted; and a submultiplicativity
        law for kernel size under parallel composition
        (\S\ref{sec:explicit}). These are classical results assembled into a
        statement about where the tractability boundary lies.
  \item For inference from executions: a formulation of semantic
        deobfuscation as MDL recovery of a factored transition model over an
        architecture-constrained family, with a proper description length; a
        lemma that transition-tensor sparsity is invariant under reshaping,
        so that factored description length, not sparsity, is the object of
        study (\S\ref{sec:factored}).
  \item A catalogue of the sources of ambiguity in factorization recovery, a
        definition of identifiability relative to a trace set, and a map from
        standard protections to the ambiguity each induces
        (\S\ref{sec:ambiguity}).
  \item Four independent falsifiable experiments, stated as multi-trace or
        active-query problems (\S\ref{sec:experiments}).
\end{enumerate}

\section{Preliminaries}
\label{sec:prelim}

A machine is a labelled transition system (LTS)
$M=(Q,\Sigma,\longrightarrow,q_0)$ with ${\longrightarrow}\subseteq
Q\times\Sigma\times Q$; $\Reach(M)$ is the set of states reachable from
$q_0$, and we identify $M$ with its reachable part. An \emph{observation
model} is a map $\pi_O:\Tr(M)\to O$ on traces; $M_1\approx_O M_2$ when the
images agree. Three models recur \cite{milner1989}: trace equivalence
$\treq$, strong and branching bisimilarity $\bisim$, and I/O equivalence.
$T$ is \emph{$O$-preserving} when $T(M)\approx_O M$; $[M]_O$ is the
equivalence class.

\begin{lemma}[observation preservation]
\label{lem:preserve}
If $T$ is $O$-preserving then $[T(M)]_O=[M]_O$. Consequently $T$ cannot
remove information required to reproduce the $O$-observable behaviour; it
may remove any information outside $O$, including information relevant to
other reverse-engineering objectives.
\end{lemma}

The lemma is immediate from the definitions and is stated only to fix what
follows from $O$-preservation and what does not. It does \emph{not} say that
reverse engineering faces no informational obstacle: under a coarse $O$,
much of what a reverse engineer wants is outside $O$ and may be gone.

Two problems must be kept apart throughout:
\[
  \boxed{\text{explicit-model contraction}\ \neq\ \text{model inference from executions}.}
\]
In the first (\S\ref{sec:explicit}) the analyst is given $M$ and asks for a
smaller $O$-equivalent system. In the second (\S\ref{sec:factored}) the
analyst holds finitely many concrete traces of $M$ and must infer a model.
Visibility of internal state settles aliasing in the second problem but says
nothing about coverage; results about the first do not transfer to the
second without an identifiability argument.

\section{Endogenous interpretation}
\label{sec:endogenous}

\subsection{One transition function, many representations}

For $Q=\mathbb{F}_2^n$ a deterministic step is $F:Q\to Q$, representable as
a transition table, as a vector of Boolean polynomials in algebraic normal
form (the representation in which mixed Boolean--arithmetic obfuscation and
its defeat live \cite{zhou2007,eyrolles2016}), or as a $2^n\times2^n$ one-hot
transition matrix. These are standard; they motivate treating the transition
function as the semantic object and instruction streams, bytecodes,
polynomials, and matrices as its representations. The one-hot matrix is also
a first example that representation cost is not intrinsic: it describes the
same $F$ exponentially more verbosely.

\subsection{Universal hosts and the three roles}
\label{sec:host}

Let $H$ be a host substrate with transition relation $\Delta_H$ on host
states $S_H$. A universal evaluator on $H$ is written
\[
  \mathcal U_H : E\times P\times Q \to Q,
\]
where $e\in E$ encodes an evaluator (an interpreter, an emulated machine
description, a specializer), $p\in P$ a program for that evaluator, and
$q\in Q$ the evaluator's own state, all three carried as data within $S_H$.
Fixing $e$ gives the evaluator's transition relation; fixing $(e,p)$ gives
the machine that program induces. The slogan ``program, interpreter, and
machine are parameters of one relation'' is true \emph{relative to $H$}:
the host's own semantics is $\Delta_H$ itself, not a parameter of
$\mathcal U_H$. This relativity is what ``endogenous'' means, and the
Futamura projections \cite{futamura1971,jones1993} are the classical
statement that the parameters of $\mathcal U_H$ can be traded against one
another by a specializer.

\subsection{Interpretation flattening}
\label{sec:flattening}

If $P$ is run by interpreter $I_1$, itself run by $I_2$, the complete host
state is a reachable subset of $S_{I_2}\times S_{I_1}\times S_P$ with one
transition relation $\Delta$.

\begin{proposition}[finite interpreter flattening]
\label{prop:flatten}
Any finite tower of effective interpreters is one effective transition
system over a reachable subset of the product of their states, and
simulations between adjacent levels compose.
\end{proposition}

Interpreter depth therefore need not imply irreducible semantic depth,
consistent with the derivation of virtual machines from interpreters
\cite{ager2003} and with specialization removing an interpretation layer
\cite{jones1993}; Giacobazzi, Jones, and Mastroeni run the same construction
backwards, obtaining obfuscators by specializing distorted interpreters
\cite{giacobazzi2012}. The flattened $\Delta$ is not a synchronous product of
level relations, which matters for composition (\S\ref{sec:composition}).

\subsection{Endogenous machines}

An executable image $B$ induces reusable state transformers
$\Gamma(B)=\{g_1,\ldots,g_m\}$ and, with a sequencing convention, a
transformation monoid $\langle\Gamma(B),\circ,\id\rangle$; a chain of gadget
addresses denotes a composition. Roemer et al.\ construct Turing-complete
gadget sets and compile a high-level language to chains \cite{roemer2012}.
Virtualization obfuscators \cite{rolles2009,kinder2012} are the same
construction with handlers for gadgets and a dispatcher for the sequencing
convention. We call such a construction an \emph{endogenous machine}: the
substrate's own transformers form the instruction basis of a further
language on the same substrate.

\section{Diffusion and contraction}
\label{sec:diffusion}

An original transition $q_i\to q_j$ may be realized in a transformed system
as a path $r_0\Rightarrow\cdots\Rightarrow r_k$, related by an abstraction
$\alpha:Q'\to Q$ with $\alpha(\Delta'^{*}(r))=\Delta(\alpha(r))$.

\begin{definition}[diffusion, contraction]
An $O$-preserving $T$ is a \emph{diffusion} with respect to a representation
cost $\kappa$ when $\kappa(T(M))>\kappa(M)$. A \emph{contraction} is a
quotient, projection, synthesis, or refactorization $C:M'\mapsto\widehat M$
with $\widehat M\approx_O M'$ and $\kappa(\widehat M)<\kappa(M')$.
\end{definition}

Obfuscation replaces a factorization $F=g_k\circ\cdots\circ g_1$ by
$F=h_m\circ\cdots\circ h_1$ with $m\gg k$, possibly over an extended state
space, preserving the induced computation while making the factorization
harder to recognize. Existing deobfuscation work exploits instances of this:
emulation-based simplification \cite{yadegari2015}, synthesis of small
equivalents from I/O samples \cite{blazytko2017,david2020}, trace-informed
control-flow synthesis \cite{mariano2024}, and abstract-interpretation models
of what obfuscation hides \cite{dallapreda2009}. The question is whether
these admit a common quantitative model. Under composition the
$O$-preserving transformations form a monoid; a diffusion and a contraction
are typically a section/retraction pair ($C\circ D=\id$ on the semantic
side, $D\circ C\neq\id$ on the richer side), and a canonical normalization
$N$ with $N^2=N$, $N\circ T=N$ is expected only property-relative or on
restricted classes. \S\ref{sec:explicit} exhibits one.

\paragraph{Execution witness.}
If $P(x)\Down y$ and $T$ preserves terminating behaviour then
$T(P)(x)\Down y$, so a finite target trace witnesses the result. The
elementary consequence is the separation of a per-step execution cost
$\Cexec$ from a recognition cost $\Crec$; obfuscation may inflate the latter
while the former stays linear in trace length. The two are not directly
comparable quantities and we make no complexity-class claim from their
separation alone. The witness argument applies to the artifact as executed;
environment-keyed code \cite{sharif2008}, split execution, and hardware-bound
execution place part of the computation outside the artifact, and the
substrate must then be taken to include the environment.

\section{Explicit-model contraction}
\label{sec:explicit}

Here the analyst is given a finite LTS $M$ explicitly.

\begin{theorem}[canonical kernel]
\label{thm:kernel}
Let $\bisim$ be branching bisimilarity and $\kernel{M}=M/{\bisim}$. Then
$\kernel{M}\bisim M$; $\kernel{M}$ has the fewest states of any LTS branching
bisimilar to $M$ and is unique up to isomorphism among minimal ones; and
$\kernel{M}$ is computable in $O(m\log n)$ for $n=|\Reach(M)|$, $m$
transitions \cite{groote1990,groote2017} (for strong bisimilarity,
\cite{paige1987,kanellakis1990}).
\end{theorem}

Branching rather than strong bisimilarity is required because inserting an
internal step changes the strong bisimulation class; branching bisimilarity
absorbs inserted internal computation while preserving branching structure
visible in $\Sigma$.

\begin{definition}[diffusion ratio]
$\delta(M)=|\Reach(M)|/|\kernel{M}|\ge1$; for $T$ with $T(M)\bisim M$,
$\delta(T;M)=\delta(T(M))/\delta(M)$.
\end{definition}

$\delta$ is invariant under the choice of abstraction onto a minimal
bisimilar system, and $N(M)=\kernel{M}$ is a canonical normalization in the
sense of \S\ref{sec:diffusion}. ``Computable in $O(m\log n)$'' is relative to
the explicit reachable state count, which is exponential in state bits; the
theorem locates where difficulty is not, it does not say contraction is
cheap.

\begin{proposition}[deterministic hypothesis class]
\label{prop:det}
If $M_1$ and $M_2$ are deterministic then $M_1\treq M_2$ iff $M_1\bisim M_2$.
Consequently the smallest \emph{deterministic} LTS trace-equivalent to a
deterministic $M$ is $\kernel{M}$, computable in $O(n\log n)$
\cite{hopcroft1971}.
\end{proposition}

\begin{proposition}[nondeterministic contractions restore hardness]
\label{prop:nondet}
For deterministic $M$ and $k$, deciding whether some LTS with at most $k$
states (nondeterminism allowed) is trace-equivalent to $M$ is
PSPACE-complete \cite{jiang1993}; the minimal such LTS is in general not
unique and may be exponentially smaller than $\kernel{M}$. Trace languages
here are prefix-closed, and minimization remains hard in that setting.
\end{proposition}

Propositions~\ref{prop:det} and~\ref{prop:nondet} concern the same $M$. What
changes is the \emph{hypothesis class} the analyst admits for the
contraction: canonical and cheap within deterministic contractions, hard as
soon as nondeterministic ones are allowed, even though execution itself is
deterministic. Determinism of execution does not by itself make contraction
easy; restricting the class of contractions does.

\begin{theorem}[hardness under traces]
\label{thm:pspace}
For finite nondeterministic LTSs, deciding $\treq$ is PSPACE-complete
\cite{meyer1972,kanellakis1990}, and minimal trace-equivalent LTSs are neither
unique nor efficiently computable unless $\mathrm{P}=\mathrm{PSPACE}$.
\end{theorem}

Nondeterminism of the \emph{observed} system arises when the observation
model hides state: erasing registers, memory, or internal labels merges
distinct concrete states under one observable and the projection may become
nondeterministic. Together with Proposition~\ref{prop:nondet}, the boundary
between the tractable and intractable regimes is set on both sides by
choices the analyst makes---what to observe and what contractions to
admit---not by the transformation. This is the explicit-model form of the
constraint-bias conjecture: the substrate keeps the concrete system
deterministic; hardness enters when that structure is discarded or when the
analyst's hypothesis class outruns it.

\subsection{Composition}
\label{sec:composition}

\begin{proposition}[submultiplicativity of kernel size]
\label{prop:submult}
For synchronous parallel composition $\otimes$ \cite{milner1989},
$|\kernel{(M_1\otimes M_2)}|\le|\kernel{M_1}|\cdot|\kernel{M_2}|$, since
bisimilarity is a congruence for $\otimes$ and $\kernel{\cdot}$ is minimal.
The inequality can be strict. When the product is fully reachable this gives
$\delta(M_1\otimes M_2)\ge\delta(M_1)\delta(M_2)$, i.e.\ $\delta$ is
\emph{supermultiplicative} and $\log\delta$ superadditive in that case.
\end{proposition}

The flattened relation of an interpreter tower is not a synchronous product,
so no bound for towers follows; whether one exists is open
(\S\ref{sec:open}).

\section{Inference from executions: factored transition models}
\label{sec:factored}

Here the analyst is not given $M$. They hold a finite set of concrete traces
$\T=\{\tau_1,\ldots,\tau_r\}$, each $\tau=(q_0,\ldots,q_m)$ a sequence of
full host states from an emulator or hardware tracer, and must infer a
model. Visibility of the full state means no two distinct visited states are
confused (aliasing is solved); it does not mean every state or transition has
been visited (coverage is not).

\subsection{The transition tensor and the invariance of sparsity}

Write $U\in\{0,1\}^{2^n\times2^n}$, $U[q,q']=1$ iff $q\to q'$, and consider
reindexings of $U$ by bit permutations, regroupings of bits into factors, and
per-factor recodings.

\begin{lemma}[sparsity is shape-invariant]
\label{lem:nnz}
For a total deterministic $F$, $\nnz(U)=2^n$, and every permutation,
regrouping, reshaping, or bijective recoding of $U$ preserves $\nnz$.
\end{lemma}

Sparsity of the transition tensor therefore cannot measure diffusion and is
not what obfuscation changes. What changes is the \emph{factored
description}: how compactly $F$ can be written as a collection of local
functions on small parent sets. That is the object of the rest of this
section, and it is exactly the object of factored transition models in
planning and probabilistic inference \cite{boutilier1999,dean1989}, where
enormous transition spaces are represented by dynamic Bayesian networks with
small parent sets. The tensor picture is retained only as intuition.

\subsection{Factorizations}

\begin{definition}[factorization]
\label{def:fact}
A \emph{factorization} $\sigma=(\pi,\varphi,\rho,D)$ of $F$ consists of
\begin{enumerate}[label=(\roman*)]
  \item a partition $\pi=\{B_1,\ldots,B_k\}$ of the $n$ state bits into
        factors;
  \item a recoding $\varphi=(\varphi_j)$, each $\varphi_j$ a bijection on
        $\mathbb{F}_2^{|B_j|}$ from an admissible family $\Phi$;
  \item a role $\rho(B_j)\in\{\mathrm{fixed},\mathrm{state}\}$ per factor;
  \item a dependency hypergraph $D=(D_j)$: for each state factor $B_j$, a
        set of parents, each either a whole factor (\emph{direct}) or an
        \emph{indexed window} $B[S]_w$ of width $w$ into a factor $B$ at an
        address held in a state factor $S$.
\end{enumerate}
$\sigma$ \emph{represents} $F$ if there are local functions
$f_j:\mathbb{F}_2^{\width(D_j)}\to\mathbb{F}_2^{|B_j|}$ with
$\varphi_j(F(q)|_{B_j})=f_j(\varphi(q)|_{D_j})$ for all reachable $q$, and
fixed factors are constant on reachable states.
\end{definition}

The ``interpreter'' of a factorization is a derived notion: a factor whose
value selects the window another factor reads (a dispatcher, a virtual
program counter). Indexed windows are what make an interpreter expressible:
the handler chosen by the opcode at the virtual PC depends on
$\text{bytecode}[\text{vpc}]_8$, not on the whole array.

\begin{proposition}[slicing as extensional specialization]
\label{prop:roles}
Fixing the fixed factors of a factorization at a value $p$ yields the
transition relation of a machine on the state factors alone. If a state
factor is constant on every execution of that machine, it may be moved into
the fixed factors and the slice taken again. This is the \emph{extensional
analogue} of fixing the static argument in the first Futamura projection:
it exhibits the specialized semantics but does not construct a residual
program, which requires a specializer as in \S\ref{sec:host}.
\end{proposition}

\subsection{Consistency with a trace set}

\begin{definition}[consistency]
\label{def:consistent}
$\sigma\models\T$ if (a) every fixed factor is constant on all states in
$\T$ and is a parent of some state factor; and (b) for each state factor
$B_j$, the observed pairs
$\big(\varphi(q_i)|_{D_j},\varphi_j(q_{i+1})|_{B_j}\big)$ over all
consecutive $(q_i,q_{i+1})$ in $\T$ define a partial function, i.e.\ contain
no conflicting entries.
\end{definition}

Checking consistency is one pass over $\T$ with hash lookups. A trace set can
only \emph{refute} a factorization: consistency establishes that the local
tables observed so far are functional, not that they are complete or that
the parent sets are minimal.

\subsection{Description length}

\begin{definition}[description length]
\label{def:mdl}
For $\sigma\models\T$,
\[
  \kappa(\sigma;\T)\;=\;\ell(\pi,\rho,D)\;+\;\ell(\varphi)\;+\;
  \sum_{j:\ \rho(B_j)=\mathrm{state}}\ell(f_j)\;+\;
  \sum_{j:\ \rho(B_j)=\mathrm{fixed}}|B_j|\;+\;\ell(\T\mid\sigma),
\]
where $\ell(\pi,\rho,D)$ is the code length of the structure under a fixed
prefix code over $\adm$, $\ell(\varphi)$ that of the recodings under a fixed
encoding of $\Phi$, $\ell(f_j)$ the length of the local function (as a table,
$|B_j|\cdot2^{\width(D_j)}$; or as a decision diagram or ANF when smaller),
and $\ell(\T\mid\sigma)$ the length of the traces given the model, which is
zero for the deterministic, exactly consistent case except for the initial
states.
\end{definition}

Charging for $\pi$, $\rho$, and $D$ is essential: they are the objects
searched over, and a cost that omits them lets the optimizer acquire
structural complexity for free. Indexed windows are charged for the window
width, not for the array read, since otherwise the cost model penalizes by
$2^{|\text{bytecode}|}$ exactly the interpreter structure to be recovered.

\begin{definition}[semantic deobfuscation as factorization recovery]
\label{def:fit}
Given an admissible family $\adm$ and a trace set $\T$,
\[
  \boxed{\;
  \sigma^\star(\T)=\operatorname*{arg\,min}_{\sigma\in\adm,\ \sigma\models\T}
  \kappa(\sigma;\T).\;}
\]
\end{definition}

This is MDL structure learning \cite{rissanen1978} of a factored transition
model from trajectories. That problem is established for Boolean and
dynamical networks \cite{akutsu1999,lahdesmaki2003,friedman1998}; what is
specific here is the hypothesis class and what is recovered from it.

\begin{definition}[admissible factorizations]
\label{def:adm}
$\adm$ is generated by the substrate's own boundaries: each architectural
register and architecturally named sub-register, each flag, each stack slot,
and each contiguous memory region touched in $\T$ is a candidate atomic
factor; factors may be merged; recodings are drawn from a small family
(identity, affine over $\mathbb{F}_2$, byte permutations); parent sets are
direct factors or indexed windows whose address factor is a register or
slot. Factorizations that cut an architectural unit at a non-architectural
boundary are excluded.
\end{definition}

This is the constraint-bias conjecture made operational: the analyst does not
search over all factorizations of $F$---partitions alone are counted by the
Bell number $B_n$---but over those the machine could have executed, and the
conjecture is that $\adm$ is small enough that $\sigma^\star$ can be found
where the unrestricted problem cannot. Whether it is small enough is
empirical (\S\ref{sec:experiments}).

\paragraph{Scope.}
The conjecture is about substrate-constrained protections: virtualization,
flattening, arithmetic encoding, code reuse. It excludes cryptographic
obfuscation. An indistinguishability-obfuscated program with an embedded
pseudorandom-function key \cite{garg2013} preserves $[P]_O$ yet admits no
efficient recovery of a small equivalent without breaking the PRF; an
obfuscated point function is a password hash. There a compact factorization
exists but is hidden by a hardness assumption, not by the combinatorics of
$\adm$, and factorization recovery must fail.

\paragraph{What visibility buys.}
Because each visited state is fully visible, the sub-LTS induced by $\T$ is
recovered exactly: no two distinct visited states are merged by observation,
and Proposition~\ref{prop:det} applies to that sub-LTS within a deterministic
hypothesis class. What visibility does not buy is coverage: from $\T$ the
analyst knows one successor for each visited state and nothing about
untaken branches. Every claim in this section is therefore relative to $\T$,
and the gap between $\T$ and $M$ is the subject of the next section.

\section{Ambiguity}
\label{sec:ambiguity}

The set of factorizations consistent with $\T$ is a version space. This
section says what makes it large, how to count it, what obfuscation does to
it, and when it collapses.

\subsection{Counting}

Two factorizations are \emph{gauge-equivalent}, $\sigma\cong\sigma'$, if they
differ only by relabelling factors, permuting bits within a factor together
with the corresponding change of $\varphi$, or composing $\varphi$ with a
bijection that leaves every local function's table unchanged up to
relabelling. Gauge-equivalent factorizations represent the same model and
must not be counted separately.

\begin{definition}[version space, ambiguity]
\label{def:ambiguity}
$V(\T)=\{[\sigma]_{\cong}:\sigma\in\adm,\ \sigma\models\T\}$ and
$A(\T)=|V(\T)|$. For a prefix length $k$ applied to every trace,
$A_\T(k)=|V(\T_{\le k})|$, non-increasing in $k$.
\end{definition}

$A$ is a version-space size. It is \emph{not} a lower bound on the running
time of any recovery procedure: SAT, constraint propagation, and
branch-and-bound eliminate exponentially many members at once, and a runtime
bound would require a query or comparison model that this note does not
supply. $A$ measures how underdetermined the model is by the data, which is
a property of the data and the hypothesis class, not of the algorithm.

\begin{definition}[identifiability]
\label{def:ident}
$F$ is \emph{identifiable from $\T$ within $\adm$} if $V(\T)$ contains a
unique class of minimal $\kappa$, and \emph{recoverable} if in addition that
class represents $F$ on all of $\Reach(M)$, not only on the states in $\T$.
\end{definition}

Identifiability is about the data singling out a model; recoverability adds
that the model is right where the data did not look. The second requires an
argument about coverage that consistency alone cannot give.

\subsection{Sources of ambiguity}

The following are the ways $V(\T)$ can contain more than one class. They are
distinct, they compound, and different protections exploit different ones.

\paragraph{(A1) Gauge ambiguity.}
Relabellings and recodings that leave the model unchanged. Removed by
counting classes rather than factorizations. Not a real ambiguity, but a
common source of over-counting in a naive $A$.

\paragraph{(A2) Coverage ambiguity.}
States and transitions not in $\T$. Every factorization consistent with $\T$
is free on the unvisited part, so $V(\T)$ contains classes that agree on
$\T$ and disagree elsewhere. This is the ambiguity that traces alone cannot
remove and that additional traces or active queries (chosen inputs, forced
branches) reduce. It is the reason a single trace cannot recover a
control-flow graph whose branches it did not take.

\paragraph{(A3) Dependency ambiguity.}
Several parent sets $D_j$ consistent with $\T$ for the same factor. On short
traces a factor may appear to depend on bits that are merely correlated with
its true parents; conversely a true parent whose value never varied in $\T$
is invisible. MDL prefers the smallest consistent parent set, which is
correct only if $\T$ is long enough to have exercised the true parents.

\paragraph{(A4) Role ambiguity.}
A factor that is constant on $\T$ may be fixed (program, table) or a state
factor that happened not to change. Requiring fixed factors to be
\emph{read} (Definition~\ref{def:consistent}) removes untouched constants
but not a variable that was read and never written in $\T$. Role ambiguity
is resolved by a write in some trace, or left open with the fixed reading
preferred by $\kappa$.

\paragraph{(A5) Granularity ambiguity.}
Two factorizations with equal $\kappa$ that differ by merging or splitting
factors. MDL ties are real ties: the data do not distinguish a two-factor
model from a merged one when the local functions have the same total length.
Ties should be reported, not broken arbitrarily.

\paragraph{(A6) Level ambiguity.}
In an interpreter tower, a bytecode array may be read as the program of the
interpreter (fixed) or as data of the flattened machine (state that is read
and never written). Both are consistent and both represent $F$; they differ
in which slice of $\mathcal U_H$ the analyst is looking at
(\S\ref{sec:host}). This is not an error but the endogenous-interpretation
phenomenon itself: the level at which a computation is ``the program'' is a
choice of fixed parameters, and $V(\T)$ contains one class per admissible
choice. $\kappa$ prefers the choice with the shortest total description,
which for a virtualized function is the one that isolates the interpreter.

\paragraph{(A7) Observation-model ambiguity.}
Different $O$ give different equivalence classes and hence different
$\sigma^\star$. A factorization minimal under I/O equivalence may drop
factors that a finer $O$ (timing, memory traffic) requires. This ambiguity
is not in $V(\T)$ but above it: it is the analyst's choice of what counts
as behaviour, and it must be fixed before $V$ is defined.

\subsection{What protections do to the version space}

Each standard protection increases one or more of these ambiguities, and
this gives a more discriminating description of a protection than
``increases complexity.''

\begin{center}
\begin{tabular}{@{}lll@{}}
\toprule
Protection & Primary ambiguity & Mechanism \\
\midrule
Virtualization & A6, A4 & program becomes read-only data of a new level \\
Control-flow flattening & A3 & next-state depends on a dispatcher variable \\
Mixed Boolean--arithmetic & A1$\to$A3 & recoding makes true parents look wider \\
Opaque predicates & A2 & branches never taken remain consistent \\
Code reuse (ROP) & A4, A6 & stack becomes the program of an induced machine \\
Junk / dead code & A3, A5 & spurious parents and factors with no effect \\
Environment keying & A2 & coverage impossible without the key \\
\bottomrule
\end{tabular}
\end{center}

The map is a hypothesis about mechanism, not a result; it predicts, for
instance, that flattening should be undone by longer traces (A3 shrinks with
data) while opaque predicates should not (A2 does not shrink without active
queries), and that virtualization should be undone by a change of level
rather than by more data. Those predictions are testable.

\subsection{When the version space collapses}

$V(\T)$ collapses to one class when $\T$ is rich enough that every true
parent has varied, every reachable branch has been taken, every factor has
been written if it is state, and $\adm$ excludes the gauge alternatives.
The sample complexity of that collapse---how many traces, of what length,
chosen how---is the identifiability question proper, and it is the theorem
this framework should eventually contain. For the deterministic finite-state
case there is an active-learning baseline: Angluin's $L^\ast$
\cite{angluin1987} learns the minimal automaton with polynomially many
membership and equivalence queries, and a factored-model analogue with
architecture-constrained hypotheses would be the natural target. We do not
have that result; we state it as the open problem the rest of the paper
points at.

\section{Proposed evaluation}
\label{sec:experiments}

The formulation predicts, for each protection, which factorization
$\sigma^\star$ should recover and which ambiguity must be reduced to get
there. The experiments below test that on independent axes: partition (E1,
E2), recoding (E3), role (E4). Each is stated as a multi-trace or
active-query problem, with a falsification criterion. Common pipeline:
compile a small function; protect it; record full-state traces in an
emulator for a chosen input set; enumerate $\adm$; compute $\sigma^\star(\T)$
by branch-and-bound on $\kappa$; compare with the ground-truth structure of
the unprotected build; report $A_\T(k)$ for protected and clean builds.

\paragraph{E1: Virtualization (partition, level).}
Tigress \texttt{Virtualize} \cite{tigress} with a table dispatcher; inputs
chosen to exercise every handler at least once (checked against the
unprotected build). \emph{Prediction.} $\sigma^\star$ places the bytecode in
a fixed factor, the virtual PC and virtual registers in state factors, and
the native handler PC in a state factor with the single parent
$\text{bytecode}[\text{vpc}]_8$; slicing away the fixed factors yields a
machine on the virtual registers whose kernel is branching bisimilar to the
kernel of the unprotected function on the covered states. \emph{Falsified
if} the MDL-optimal admissible factorization does not isolate the handler
table, or if a handler exercised in $\T$ is not recovered. \emph{Baseline.}
Syntia \cite{blazytko2017}, QSynth \cite{david2020}.

\paragraph{E2: Control-flow flattening (dependency).}
Tigress \texttt{Flatten} or O-LLVM \texttt{-fla} \cite{ollvm}; inputs chosen
to cover every original edge, or active queries forcing the untaken ones.
\emph{Prediction.} $\sigma^\star$ isolates the dispatcher variable as a state
factor whose local function, on the covered edges, has the original
control-flow graph as its graph. \emph{Falsified if} MDL merges the
dispatcher with data, or if the recovered graph is not isomorphic to the
covered part of the original CFG. Also measure how $A_\T(k)$ falls with
added traces (the A3 prediction). \emph{Baseline.} Mariano et al.\
\cite{mariano2024}.

\paragraph{E3: Mixed Boolean--arithmetic (recoding).}
Tigress \texttt{EncodeArithmetic} or the identities of Zhou et al.\
\cite{zhou2007}; $\pi$ trivial, search over $\varphi\in\Phi$ affine plus
bit-slicing. \emph{Prediction.} $\sigma^\star$ is a recoding under which the
local function's ANF has degree $\le2$, recovering the original operator.
\emph{Falsified if} no admissible recoding lowers the degree, which would
mean the identity is not a change of basis within $\Phi$---a finding about
$\Phi$, to be reported as such. \emph{Baseline.} Eyrolles et al.\
\cite{eyrolles2016}.

\paragraph{E4: Return-oriented chain (role, level).}
A small function compiled to a chain against a fixed binary
\cite{roemer2012}. \emph{Prediction.} $\sigma^\star$ assigns the chain region
the fixed role, the stack pointer a state role with itself as parent, and
the instruction pointer a state role with parent $\text{stack}[\text{sp}]_{64}$,
i.e.\ the chain is recognized as a program and gadget addresses as its
opcodes. \emph{Falsified if} the chain region is not assigned the fixed role
or the sliced machine is not bisimilar to the original on covered states.

\paragraph{Failure of the approach.}
The approach is refuted, not merely incomplete, if in any of E1--E4 the
MDL-optimal factorization within $\adm$ fails to coincide with ground truth
on the covered part while a factorization outside $\adm$ does: that would
falsify the constraint-bias conjecture directly. Failure due to coverage
(A2) is not a refutation of the approach but a measurement of sample
complexity, and must be reported separately.

\section{Related work}
\label{sec:related}

\paragraph{Cryptographic obfuscation.}
Barak et al.\ \cite{barak2001} construct function families with a predicate
efficiently recoverable from any implementation but not from oracle access,
showing virtual-black-box obfuscation impossible in general.
Indistinguishability obfuscation \cite{garg2013} guarantees computational
indistinguishability of obfuscations of functionally equivalent circuits;
best-possible obfuscation \cite{goldwasser2007} reveals no more than any
equivalent program does. Lemma~\ref{lem:preserve} is consistent with these
results and says nothing beyond them; cryptographic obfuscation is outside
the scope of \S\ref{sec:factored}.

\paragraph{Semantics-based obfuscation and deobfuscation.}
Collberg et al.\ \cite{collberg1997} introduced the potency/resilience/cost
vocabulary. Dalla Preda and Giacobazzi model obfuscation as incompleteness of
an abstract interpretation \cite{dallapreda2009,cousot1977}; Giacobazzi,
Jones, and Mastroeni obtain obfuscators by specializing distorted
interpreters \cite{giacobazzi2012}, the closest prior work to
\S\ref{sec:flattening}. Schrittwieser et al.\ survey the arms race
\cite{schrittwieser2016}; Ollivier et al.\ give an analyzer-relative potency
measure \cite{ollivier2019}. On the recovery side, Rolles \cite{rolles2009}
and Kinder \cite{kinder2012} treat VM obfuscators as machines to be
modelled; Yadegari et al.\ \cite{yadegari2015} simplify from traces; Syntia
\cite{blazytko2017} and QSynth \cite{david2020} synthesize from I/O; Mariano
et al.\ \cite{mariano2024} synthesize control flow; Eyrolles et al.\
\cite{eyrolles2016} normalize MBA. Each is a contraction under some $O$.

\paragraph{Factored transition models and structure learning.}
Dynamic Bayesian networks \cite{dean1989} and factored MDPs
\cite{boutilier1999} represent large transition systems by local functions
with small parent sets; learning their structure from time series is
established for DBNs \cite{friedman1998} and for Boolean networks
\cite{akutsu1999,lahdesmaki2003}, with MDL as a standard criterion
\cite{rissanen1978}. Angluin's $L^\ast$ \cite{angluin1987} learns minimal
automata from queries. \S\ref{sec:factored} is an instance of this
literature; the claimed novelty is the architecture-constrained hypothesis
class, the recovery of program and interpreter roles as structure, and the
connection to protections on binaries.

\paragraph{Interpreters, partial evaluation, code reuse.}
Futamura \cite{futamura1971}, Jones et al.\ \cite{jones1993}, Ager et al.\
\cite{ager2003}, and Roemer et al.\ \cite{roemer2012} supply the established
ingredients of endogenous interpretation. The semantic gap of
virtual-machine introspection \cite{garfinkel2003,jain2014} is a recognition
problem over a complete artifact and is in that respect closer to this note
than the decompilation usage.

\paragraph{Equivalence and minimization.}
Paige--Tarjan \cite{paige1987}, Kanellakis--Smolka \cite{kanellakis1990},
Groote--Vaandrager \cite{groote1990} and Groote et al.\ \cite{groote2017}
for bisimulation quotients; Hopcroft \cite{hopcroft1971} for DFA
minimization; Meyer--Stockmeyer \cite{meyer1972} and Jiang--Ravikumar
\cite{jiang1993} for the PSPACE results; Milner \cite{milner1989} for the
spectrum of equivalences.

\section{Established, proposed, open}
\label{sec:open}

\paragraph{Established.}
Universal interpretation and specialization; simulation and bisimulation;
derivation of VMs from interpreters; $O$-preserving transformations;
abstract-interpretation models of obfuscation; semantic deobfuscation and
synthesis; code-reuse computation; finite-state minimization and its
complexity; factored transition models and MDL structure learning.

\paragraph{Proposed.}
Endogenous interpretation as a viewpoint; semantic diffusion in place of the
gap metaphor for $O$-semantics; the tractability boundary of explicit-model
contraction as a function of observation model and hypothesis class;
semantic deobfuscation as architecture-constrained factorization recovery
with a proper description length; the ambiguity catalogue and the
protection-to-ambiguity map.

\paragraph{Open.}
\begin{enumerate}
  \item Identifiability: a sample-complexity result for factorization
        recovery within $\adm$, in a passive or active model, in the style
        of $L^\ast$ for the factored, architecture-constrained class. This is
        the theorem the framework needs.
  \item Whether a composition bound holds for interpreter towers, where the
        flattened relation is not a synchronous product.
  \item Where in the equivalence spectrum between branching bisimilarity and
        trace equivalence the cost of contraction jumps, and which
        deobfuscation technique's observation model sits where.
  \item Whether $\adm$ for a real ISA admits exact search or only greedy
        search, and which protections defeat the greedy variant.
  \item Whether the protection-to-ambiguity map predicts, on E1--E4, which
        interventions (more traces, active queries, change of level) reduce
        $A_\T$ for which protection.
\end{enumerate}

\paragraph{Thesis.}
\begin{quote}
\textbf{Under a chosen observation model, an $O$-preserving transformation
redistributes rather than removes the computation that produces the
observable behaviour. For explicitly given finite-state realizations the
cost of contracting it is set by the observation model and the admitted
hypothesis class, not by the transformation. For realizations known only
through executions, semantic deobfuscation is the recovery of a compact
factored transition model over the factorizations the substrate could have
executed; protections act by enlarging the version space of consistent
factorizations, and the kind of ambiguity they induce determines what
reduces it.}
\end{quote}


\end{document}